\documentclass[sigconf]{acmart}
\renewcommand\footnotetextcopyrightpermission[1]{} 
\usepackage{dsfont}
\usepackage{booktabs}
\usepackage{tikz}
\usepackage{float}
\usetikzlibrary{external}
\usepackage{textcomp}
\usepackage[utf8]{inputenc}
\usepackage{enumitem}

\usepackage{makecell}
\usepackage{threeparttable}
\usepackage{longtable}
\usepackage{multirow}
\usepackage{tabularx}
\usepackage{tabulary}
\usepackage{array}
\newcommand{\PreserveBackslash}[1]{\let\temp=\\#1\let\\=\temp}

\newcolumntype{C}[1]{>{\PreserveBackslash\centering}p{#1}}
\newcolumntype{R}[1]{>{\PreserveBackslash\raggedleft}p{#1}}
\newcolumntype{L}[1]{>{\PreserveBackslash\raggedright}p{#1}}
\usepackage{amssymb}
\usepackage{eqnarray}
\usepackage{amsmath}
\usepackage{graphicx}
\usepackage{subfig}
\usepackage{caption}
\usepackage{color}
\usepackage{url}
\usepackage{bm}

\usepackage[linesnumbered,ruled,vlined]{algorithm2e}
\usepackage{hyperref}

\def\BibTeX{{\rm B\kern-.05em{\sc i\kern-.025em b}\kern-.08emT\kern-.1667em\lower.7ex\hbox{E}\kern-.125emX}}

\copyrightyear{2019}
\acmYear{2019}
\acmConference[MM '19]{Proceedings of the 27th ACM International Conference on
Multimedia}{October 21--25, 2019}{Nice, France}
\acmBooktitle{Proceedings of the 27th ACM International Conference on Multimedia (MM '19),October 21--25, 2019, Nice, France}
\acmPrice{15.00}
\acmDOI{10.1145/3343031.3350953}
\acmISBN{978-1-4503-6889-6/19/10}

\begin{document}
%
\title{User Diverse Preference Modeling by Multimodal Attentive Metric Learning}
\author{Fan Liu$^1$, Zhiyong Cheng$^2$, Changchang Sun$^1$, Yinglong Wang$^2$, Liqiang Nie$^1$, Mohan Kankanhalli$^3$}

\affiliation{$1.$School of Computer Science and Technology, Shandong University}
\affiliation{$2.$Shandong Computer Science Center (National Supercomputer Center in Jinan), \\ Qilu University of Technology (Shandong Academy of Sciences)}
\affiliation{$3.$School of Computing, National University of Singapore}

\email{{liufancs, jason.zy.cheng, nieliqiang }@gmail.com, mohan@comp.nus.edu.sg}
\thanks{Corresponding author: Zhiyong Cheng and Liqiang Nie}

%
%
\begin{abstract}
Most existing recommender systems represent a user's preference with a feature vector, which is assumed to be fixed when predicting this user's preferences for different items.  However, the same vector cannot accurately capture a user's varying preferences on all items, especially when considering  the diverse characteristics of various items.  To tackle this problem, in this paper, we propose a novel Multimodal Attentive Metric Learning (MAML) method to model user diverse preferences for various items. In particular, for each user-item pair, we propose an attention neural network, which exploits the item's multimodal features to estimate the user's special attention to different aspects of this item. The obtained attention is then integrated into a metric-based learning method to predict the user preference on this item. The advantage of metric learning is that it can naturally overcome the problem of dot product similarity, which is adopted by matrix factorization (MF) based recommendation models but does not satisfy the triangle inequality property.  In addition, it is worth mentioning that the attention mechanism cannot only help model user's diverse preferences towards different items, but also overcome the geometrically restrictive problem caused by collaborative metric learning. Extensive experiments on large-scale real-world datasets show that our model can substantially outperform the state-of-the-art baselines, demonstrating the potential of modeling user diverse preference for recommendation.
\end{abstract}

%
%


%

%
\maketitle

\section{Introduction}
Recommendation has become one of the most important techniques for various online platforms, such as E-commerce (e.g., Amazon\footnote{https://www.amazon.com.}), streaming service (e.g., Youtube\footnote{https://www.youtube.com.}), and social platforms (e.g., Pinterest\footnote{https://www.pinterest.com.}).  Due to its importance, many recommendation techniques have been developed so far. Among them, matrix factorization (MF) has achieved great success~\cite{koren2009matrix,bell2007lessons}. Given the user-item interaction matrix, this method maps all users and items into a latent feature space, in which each user and item is presented as a feature vector. The preference of a user ($u$) for an item ($i$) is predicted based on the dot product of the feature vectors, i.e., $\bm{p}_u^T\bm{q}_i$. Many recommendation methods have been proposed based on this idea, like WRMF~\cite{hu2008collaborative}, BPR~\cite{rendle2009bpr}, and  PMF~\cite{salakhutdinov2007probabilistic}. In recent years, the matrix factorization for recommendation has been greatly advanced with the development of deep neural network (DNN) techniques. A set of DNN based matrix factorization methods has been proposed, such as the neural collaborative filtering (NCF)~\cite{he2017neural} and Deep MF~\cite{xue2017deep}.

 The success of MF is largely attributed to its simple and effective idea of exploiting the user-item interaction data to mine user preferences. However, merely relying on the interaction information also leads to some drawbacks, like 1) the incapability of modeling the  fine-grained user preferences at the feature-level or aspect-level; and 2) performance degradation problem when the interaction data for items or users are insufficient~\cite{cheng2018aspect}. To alleviate those limitations, various types of side information have been incorporated in MF, such as item images~\cite{wang2017your,kalantidis2013getting,cheng2019mmalfm}, metadata~\cite{cheng2016wide,chen2018attention}, and user reviews~\cite{mcauley2013hidden,tang2015user,cheng2018aspect}. The intuition is that side information contains item cues which adds value to user preference modeling. Among the side information, user reviews contain users' opinions on different aspects of items and thus have been widely exploited to model  fine-grained user preferences~\cite{chin2018anr,cheng20183ncf,chen2018neural,cheng2018aspect}. Besides, item images can help capture user preferences on the visual appearance of items, which have been extensively explored in fashion recommendation~\cite{he2016vbpr,mcauley2015image,song2017neurostylist}. Different types of side information are considered to be complementary to each other. With the advancement of representation learning~\cite{bengio2013representation,Wang:2012},  several methods have been proposed to exploit multi-source information for recommendation, such as  the Collaborative Knowledge based Embedding (CKE)~\cite{zhang2016collaborative} method using structure information, textual and visual knowledge, as well as the Joint Representation Learning (JRL) method~\cite{zhang2017joint} exploiting ratings, reviews and item images.

Although great progress has been achieved thus far, most existing recommender systems have not well considered to characterize user varying preferences for different items. It is common that the user preference on the various aspects of items is different. For example,  a user will value the ``plot" more for a suspense movie, while pays more attention to the ``special effects" for a super-hero movie. From this example, we can see that users do not treat aspects equally for different items, even when the items are of the same category. Most existing recommendation methods use \emph{the same vector to represent a user's preferences for all items}, which cannot accurately predict the diverse preferences on various items. Recently, some researchers have noticed this problem and proposed several models to tackle it. One type of methods leverages reviews to analyze user attention on different aspects of items and then integrates the results into the matrix factorization methods for recommendation, such as ALFM~\cite{cheng2018aspect}, A$^3$NCF~\cite{cheng20183ncf}, and ANR~\cite{chin2018anr}. Another type of methods varies the target item or user vector based on the vectors of those most influential items or users with respect to the target items. These recently proposed methods are based on deep neural networks, such as deep interest networks (DIN)~\cite{zhou2018deep}, memory attention-aware recommender system (MARS)~\cite{zheng2018mars}, and collaborative memory network (CMN)~\cite{ebesu2018collaborative}.

With the consideration of users' varying preferences towards different items,  the above methods have achieved better results over their baselines. They, however, still rely on the matrix factorization method with dot product for similarity prediction. An intrinsic problem of the dot product is that it is not a metric-based distance learning, as it does not satisfy the triangle inequality\footnote{It is defined as: ``the distance between two points cannot be larger than the sum of their distances from a third point''~\cite{tversky1982similarity}.}~\cite{hsieh2017collaborative,Yang2018PersonRW}. This problem limits the expressiveness of MF and hence causes MF-based methods incapable of capturing the fine-grained user preferences, resulting in sub-optimal performance~\cite{hsieh2017collaborative,zhang2018metric}. To tackle this problem, Hsieh et al.~\cite{hsieh2017collaborative} proposed a metric collaborative filtering (CML) method which learns the preferences by minimizing the distance between user and item vectors (i.e., $||\bm{p}_u  - \bm{q}_i||$) of positive interactions. However, directly applying of this distance metric is problematic~\cite{Wang2013ViewBasedDP}. Intuitively, this method tries to map each user and item pair with positive interaction to the same point in a low dimensional space. The problem is that each user and item has many interacted items and users, respectively. As a result, it is geometric inflexible for a large dataset since it tries to fit a user and all the interacted items into the same point. Tay et al.~\cite{tay2018latent} has mathematically proved this is  \emph{geometrically restrictive} and will lead to an \emph{ill-posed algebraic} system. They proposed to learn an adaptive relation vector $r_{ui}$ between the interactions of a user-item pair by minimizing $||\bm{p}_u + \bm{r}_{ui} - \bm{q}_i||$. Although those proposed metric-based learning approaches tackled  the problem of dot product, they have not well modeled user diverse preferences for various items.
\vspace{-0.05cm}

Based upon the aforementioned insights, we present a novel multimodal attentive metric learning (MAML) method by leveraging the multimodal information of items to model user diverse preferences towards different items. Our MAML model enhances the CML method by introducing a weight vector to each user-item pair. This weight vector represents the user's attention on different aspects of the target items, and thus it is unique for each user-item pair. This is achieved by a designed attention neural network which analyzes user attention on the target item by exploiting the textual and visual features of this item. The learned attention vector is then integrated with the user and item vectors to compute the user and item distance. In this way, our model can capture a user's varying preferences for different items in recommendation. It is also worth mentioning that our proposed MAML enjoys the following merits: 1) It satisfies the inequality property because of the adopted metric-based learning approach, and thus avoids the problem of the dot product similarity prediction method; And 2) it tackles the geometrically restrictive problem in CML. This is because the weight vector works as a transformation vector which projects the user and item into a distinct space (because the weight vector is unique for each user-item pair) to perform the proximate calculation via the Euclidean distance. Therefore, MAML allows a greater extent of geometric flexibility and modelling capability. Besides, we find that the standard attention mechanism is inferior to MAML because it leads to very small Euclidean distance between each dimension of the user and item vectors, which significantly diminish the prediction power of our model (see Sect.~\ref{sec:am} for detail). To address this, we adjust the attention design by enlarging the summation value of the attention weights. We conduct comprehensive experiments on three benchmark datasets to evaluate the performance of our model on the top-$n$ recommendation task. Experimental results demonstrate that our MAML outperforms the state-of-the-art MF-based and metric-based learning approaches. Beyond the improvement in accuracy, an exciting property of MAML is that it can uncover users’ varying preferences on items. 

\section{Related Work}


\textbf{Diverse Preference Modeling.} 
In recent two years, researchers have paid more attention to model the varying preferences towards different items and proposed several methods. We roughly categorize those methods into two groups. 1) Methods in the first group exploit reviews to analyze each user's attention on different aspects of the target item, and then integrate the attention weights into the matrix factorization for recommendation~\cite{cheng2018aspect,cheng20183ncf,chin2018anr,chen2018attention}. In particular, ALFM~\cite{cheng2018aspect} applies a topic model on reviews to detect the user attention, and  A$^3$NCF~\cite{cheng20183ncf} uses a neural attention network to learn the user attention from reviews. Following the same idea, Chin et al.~\cite{chin2018anr} developed an end-to-end attentive neural network based recommendation model, which leverages reviews and ratings to learn user diverse preferences on different aspects of items. The AFM method~\cite{chen2018attention} adopts a similar strategy and exploits other types of side information (e.g., item category) instead of reviews.
And 2) as to the methods in the second group, they adapt the user or item vector according to the most influential users or items, which are selected based on the target item. For example, in CMN~\cite{ebesu2018collaborative}, the target user vector is computed based on the neighbour user  vectors, where the neighbour users and their contributions to the target user vector depend on the target item. MARS~\cite{zheng2018mars} adapts the target user vector based on the neighbour item vectors of the target item. Both CMN and MARS fix the vector of the target item while adapt the user vector. In contrast, DIN~\cite{zhou2018deep} fixes the user preference vector while adapts the target item representation based on the user's previously purchased items.

All the aforementioned methods are based on the matrix factorization framework by using dot product to estimate the similarity between the target user and item. With a different strategy in this work, we developed a novel user diverse modeling method based on the metric learning method, which automatically avoids the limitation of dot product. To the best of our knowledge, this is the first work to adopt metric learning method to capture user diverse preferences on different items.

\textbf{Metric-based Learning.} Several efforts have been dedicated to incorporating metric-based learning methods into collaborative filtering. Typically, the Euclidean embedding is the most popular one. For instance, Khoshneshin et al.~\cite{khoshneshin2010collaborative} used the Euclidean embedding to model explicit feedback. Bachrach et al.~\cite{bachrach2014speeding} developed a method to transform the pre-trained dot-product space into a Euclidean space by preserving the item similarity. The most related studies are the recently proposed collaborative metric filtering (CML)~\cite{hsieh2017collaborative} and latent relational metric learning (LRML) methods~\cite{tay2018latent}. CML learns the user vector and item vector by minimizing the Euclidean distance of each user-item pair. Despite its simplicity, it can effectively encode the user preference, as well as user-user and item-item similarity, achieving very competitive performance on several datasets. In~\cite{tay2018latent}, authors pointed out that CML is geometrically restrictive and leads to an ill-posed algebraic system. This is because it tries to map each pair of user-item into the same point in the Euclidean space, and there are many neighbours for each user and item. To tackle this problem, they introduced a relational vector to connect the user and item in the embedding space. In this paper, we introduce an adaptive weight vector to transform each user-item pair into a unique space (i.e., the user-item specific attentive-aspect space) for minimization. Therefore, our method could also deal with the problem of CML. Besides, the weight vector is tailored for each pair of user-item, and thus our method can capture users' varying preferences upon items.

\textbf{Multimodal User Preference Modeling.} 
Although many recommender systems have been designed to exploit side information to enhance the recommendation performance,  most of them only exploit a single modality feature, such as textual reviews~\cite{catherine2017transnets,mcauley2013hidden,tan2016rating,cheng20183ncf,chen2018neural,cheng2018aspect,chin2018anr}, item images~\cite{kalantidis2013getting,wang2017your,he2016vbpr,mcauley2015image}, and other metadata~\cite{chen2018attention,cheng2016wide,gao2018recommendation,Nie2016Learning}. More recently,  several deep models have been proposed to model user preference by using multimodal information. TranSearch~\cite{guo2018multi} exploits both textual reviews and item images to model the user preference on products for personalized search. Zhang et al.~\cite{zhang2016collaborative} proposed a knowledge base method, which extracts the multimodal knowledge as well as unstructured textual and visual knowledge to jointly learn the latent representations of items within a collaborative filtering framework. Similarly, Zhang et al.~\cite{zhang2017joint} first extracted user and item features from ratings, reviews, and images via deep representation learning and then concatenated the multimodal features to form the joint representations of users and items for recommendation. In this work, our method is different from the above methods from two perspectives. Firstly, the above methods leverage the multimodal features to jointly learn fixed user and item representations for  recommendation, while our method exploits the multimodal features to learn user's varying attention to the aspects of different items.  Secondly, our method is based on a metric learning strategy and adopts the Euclidean distance to predict the user preference on an item, which is greatly different from the methods relying on the matrix factorization framework and the dot product for similarity estimation. 
\section{Our Proposed Model}
\subsection{Notations and Background}
\textbf{Notations.} Before describing of our model, we would like to first define some basic notations. Let $\mathcal{U}$ be the user set,  $\mathcal{I}$ be item set, and $\bm{R}$ be the user-item interaction matrix. $r_{u,i}\in \bm{R}$ indicates the interaction between the user $u\in \mathcal{U}$ and the item $i\in \mathcal{I}$. The interaction could be implicit or explicit. For implicit feedback, $r_{u,i}=1$ if the interaction between $u$ and $i$ exists; otherwise, $r_{u,i}=0$. For explicit feedback, $r_{u,i}$ usually is the rating that $u$ assigns to $i$.   Let $\mathcal{R}$ be the set of user-item  $(u,i)$ pairs whose values are non-zero and $\mathcal{R}_u$ be the set of items that have been interacted by $u$. $\bm{p}_u \in \mathds{R}^f$ and $\bm{q}_i \in \mathds{R}^f$  denote the latent feature vector for $u$ and  $i$, respectively.\footnote{In the paper, unless otherwise specified, notations in bold style denote matrices or
vectors, and the ones in normal style denote scalars.} The core of recommender systems is to learn the latent  feature vectors of users and items (i.e., $\bm{p}_u$ and $\bm{q}_i$). When recommending items to a user $u$, for each item $j \notin \mathcal{R}_u$, its recommendation score $\hat{r}_{u,j}$ is computed based on  $\bm{p}_u$ and $\bm{q}_j$.

\begin{figure*}[htp]
	\centering
	\includegraphics[width=0.8\linewidth]{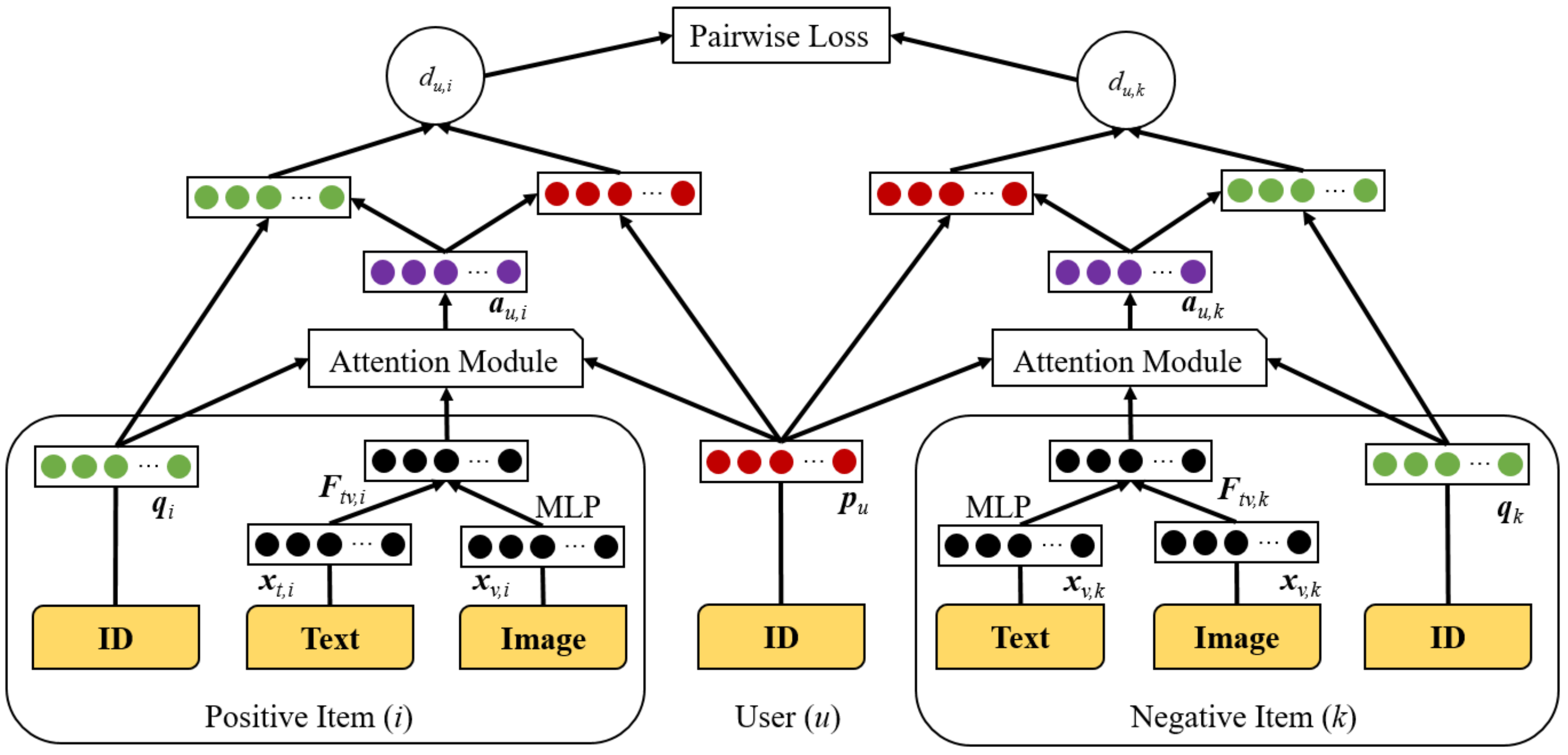}
	\caption{Overview of our MAML model.}
	\label{fig:maml}
	\vspace{-0.5cm}
\end{figure*}

\textbf{Background.} Matrix factorization (MF)~\cite{koren2009matrix} maps users and items into a shared latent feature space and estimates an unseen interaction by the dot product between the user and item vector, i.e., $\hat{r}_{u,i} = \bm{p}_u^T \bm{q}_i$. And $\bm{p}_u$ and $\bm{q}_i$ for all  $u \in \mathcal{U}$ and $i \in \mathcal{I}$ are learnt by minimizing the predicted errors between the $\hat{r}_{u,i}$ and $r_{u,i}$, namely,  $\sum_{{u,i}\in \mathcal{R}}\|\hat{r}_{u,i}-r_{u,i})\|$\footnote{Notice that here we ignore the regularization term for the simplicity of presentation.}. Due to its simplicity and superior performance, matrix factorization has become the most popular collaborative filtering method over the past decade. The original MF models were designed for rating prediction~\cite{koren2009matrix,he2017neural} and was extended to weighted regularized matrix factorization (WRMF) for implicit feedback prediction~\cite{hu2008collaborative}. As the goal of recommender system is to provide the target user with a ranking list of top few items, which are most likely to be preferred by the user, recommendation is rather a ranking problem instead of rating. In light of this, MF moves to model the relative preferences between different items and the pairwise learning approach has been widely adopted to achieve the goal~\cite{rendle2009bpr,zhang2017joint}. In pairwise learning, the user and item vectors are learnt by setting $\hat{r}_{u,i} > \hat{r}_{u,k}$ for any two pairs that satisfy $(u,i)\in \mathcal{R}$ and $(u,k)\notin \mathcal{R}$. A typical example is the bayesian personalized ranking (BPR)~\cite{rendle2009bpr}. It is a MF based method, in which $\hat{r}_{u,i}=\bm{p}_u^T \bm{q}_i$ and $\hat{r}_{u,k}=\bm{p}_u^T \bm{q}_k$ and thus $\bm{p}_u^T \bm{q}_i> \bm{p}_u^T \bm{q}_k$ (s.t. $(u,i)\in \mathcal{R}$ and $(u,k)\notin \mathcal{R}$) is used for optimization.

Despite the success of MF, it is not a metric-based learning approach as the adopted dot product similarity does not satisfy the triangle inequality property, which is critical for modeling fine-grained user preference as demonstrated in~\cite{hsieh2017collaborative,zhang2018metric}. From the metric learning perspective, since  users and items are represented as latent vectors in a shared space, the similarity between a user and an item can be estimated based on the Euclidean distance between their vectors as:
\begin{equation} \label{eq:cml}
d(u, i) = \|\bm{p}_u-\bm{q}_i\|.
\end{equation}
 Considering the pairwise learning, it is natural to derive that $d(u, i) < d(u, k)$ for $(u,i)\in \mathcal{R}$ and $(u,k)\notin \mathcal{R}$, where $i$ denotes an item that $u$ likes and $k$ denotes an item that $u$ does not like. The collaborative metric learning (CML)~\cite{hsieh2017collaborative} is designed based on this simple idea. As this method is a metric-based learning approach, it naturally avoids the limitation of dot product and achieves better performance than the MF models~\cite{hsieh2017collaborative}. Therefore, in this work, we also derive our user diverse preference modeling method based on this approach instead of MF.

\subsection{Multimodal Attentive Metric Learning}
\subsubsection{Overview}
In the aforementioned methods, they all utilize a fixed vector $\bm{p}_u$ to represent a user $u$'s preference in the feature space.  In those models which map users and items into a joint latent space for similarity estimation, they all assume that each dimension in the space stands for a type of feature or an aspect of the items. Those dimensions/aspects are expected to well describe and distinguish the preferences of different users. Based on this understanding, we argue that using the same vector $\bm{p}_u$ to predict $u$'s preferences for all items may be not optimal, because in the real scenarios, it is common that the preference of a user on the aspects of different items is varying. For example, a user who prefers the \emph{taste} and \emph{price} of a restaurant may pay more attention to the \emph{ambience} and \emph{service} of another restaurant, because the two restaurants serve for different purposes. When predicting the preference of a user $u$ towards an item $i$, those aspects of the item $i$ to which $u$ pays the most attention should dominate $u$'s preference on the item $i$.

In light of this, we propose a multimodal attentive metric learning (MAML) model. For each user-item $(u,i)$ pair, our model computes a weight vector $\bm{a}_{u,i} \in \mathds{R}^f$ to indicate the importance of $i$'s aspects for $u$.  In addition, the side information of items is exploited to estimate the weight vector, as side information conveys rich features of items, especially text reviews and item images, which are well-recognized to provide notable and complementary features of items in different aspects~\cite{zhang2017joint,cheng2019mmalfm}. We adopt the recent advancement of attention mechanism~\cite{chen2017attentive,cheng20183ncf}  to estimate the attention vector.  With the attention (weight) vector, the Euclidean distance between a user $u$ and an item $i$ in our model becomes:
\begin{equation} \label{eq:acml}
d(u, i) = \|\bm{a}_{u,i} \odot \bm{p}_u - \bm{a}_{u,i} \odot \bm{q}_i\|,
\end{equation}
where $\odot$ denotes the element-wise product between vectors.

It is worth mentioning that, with the attention vector, our model can not only accurately capture the user's varying preferences for different items, but also tackle the geometrically restrictive problem~\cite{tay2018latent} in CML. From Eq.~\ref{eq:cml}, it can be found that CML tries to fit a user and all the interacted items into the same point in the latent space, however, each item in turn has many interacted users. Therefore, it is geometrically impossible to achieve the goal. In our model, as $\bm{a}_{u,i}$ is unique for each user-item pair, it works as a transformation vector which transforms the target user and item into a new space for distance computation. Thus, our method can naturally avoid the geometrically restrictive problem in CML. An overview of our proposed modal MAML is illustrated in Fig.~\ref{fig:maml}.

We adopt the pairwise learning for optimization and the loss function is defined as:
\begin{equation} \label{eq:obj}
L_{m}\left ( d \right ) = \sum_{(u,i)\in R }\sum_{(u,k)\notin R }\omega _{ui}[m + d(u,i)^{2} - d(u,k )^{2}]_{+},
\end{equation}
where $i$ is an item that $u$ likes and $k$ is an item that $u$ does not like; $[ z ]_{+} = max( z,0)$ denotes the standard hinge loss. $\omega _{ui}$ is a ranking loss weight (described in Sect.~\ref{sec:rlw}) and $m > 0$ is the safety margin size. $d(u,i)$ and $d(u,k)$ are computed according to Eq.~\ref{eq:acml}.  In the next, we introduce how to compute the attention vector $\bm{a}_{u,i}$ for each user-item pair.

\subsubsection{Attention Mechanism} \label{sec:am}
In this section, we introduce the attention mechanism in MAML for capturing a user $u$'s specific attention $\bm{a}_{u,i}$ of an item $i$. Since text reviews and images contain rich information about user preference and item characteristic, they are used to capture $u$'s attention on the various aspects of $i$. A two-layer neural network is used to compute the attention vector:
\begin{equation}
\bm{e}_{u,i} = Tanh( \bm{W}_1 [ \bm{p}_u;\bm{q}_i;\bm{F}_{tv,i}] + \bm{b}_1 ),
\end{equation}
\begin{equation}
\bm{\hat{a}_{u,i}} = \mathbf{v}^{T}ReLU ( \bm{W}_2\bm{e}_{u,i} + \bm{b}_2 ),
\end{equation}
where $\bm{W}_1,\bm{W}_2$ and $\bm{b}_1,\bm{b}_2$ are respectively the weight matrices and bias vectors of the two layers. $\mathbf{v}$ is a vector that projects the hidden layer into an output attention weight vector. $\bm{F}_{tv,i}$ is the item feature vector which is a fusion of $i$'s textual feature and image feature (described later).  $[ \bm{p}_{u};\bm{q}_{i};\bm{F}_{tv,i} ]$ denotes the concatenation of $\bm{p_{u}}$, $\bm{q}_{i}$, and $\bm{F}_{tv,i}$. Tanh and ReLU~\cite{song2018neural,Liu:2018:AMR:3209978.3210003,Liu:2018:CML:3240508.3240549} are  used as the activation functions for the first and second layer, respectively.\footnote{We have empirically studied different combinations of the activation functions in the attention neural network, and find out that Tanh and Relu can achieve relatively better performance.}

Following the standard procedures of neural attention networks, there is a subsequent step to normalize $\bm{\hat{a}_{u,i}}$ with the softmax function, which converts the attention weights to a probabilistic distribution. Unfortunately, this standard solution does not work well in practice. This is because in our model, the attention weights are directly used to element-wise product with the Euclidean distance between $\bm{p}_u$ and $\bm{q}_i$ (see Eq.~\ref{eq:acml}). For each dimension $l$, the weighted distance is  $d_{u,i,l}=a_{u,i,l}\cdot (p_{u,l}-q_{i,l})$. After softmax normalization, the weights will be very small. For example, when the dimension $f$ is 100, the mean value of weights is only 0.01. Notice that the distance between each dimension of $\bm{p}_u$ and $\bm{q}_i$ is already quite small~\footnote{Due to the regularization $\| \bm{p}_{*}\|^{2} \leq 1$ and  $\| \bm{q}_{*} \|^{2} \leq 1$, see Sect.~\ref{sec:reg}}. With such a small weight $a_{u,i,l}$, the distance $d_{u,i,l}$ becomes even smaller. When the distances of all dimensions are quite small, the differences between different dimensions (aspects) become negligible. This will weaken  the distinguishing power of our model, resulting in performance deterioration. To alleviate this problem, we propose to enlarge the normalized weight by a factor $\alpha$. In our model, the final attention weight vector is computed as:
\begin{equation}
a_{u,i,l} = \alpha \cdot \frac{exp\left ( \hat{a}_{u,i,l} \right )}{\sum _{l=1}^{f}\hat{a}_{u,i,l}}.
\end{equation}
In experiments, we set $\alpha$ to be the dimension of the weight vector, i.e., $\alpha=f$, which turns out to work well (see Section~\ref{alpha_section}). This setting is motivated by the consideration that when the weight vector is binary, only the aspects with the weight 1 take effects on the final decision, and the extremely case is that all the aspects are of the same important.
In the next, we introduce how to obtain the fused item feature vector $\bm{F}_{tv,i}$.

\textbf{Item Features.}
For each item, its textual and visual features are extracted from its associated reviews and images.\footnote{In this work, each item is associated with one image.}  The text feature $\bm{F}_{t,i}$ is extracted by the PV-DM \cite{le2014icml} model, which learns continuous distributed vector representations for textual documents
in an unsupervised way. It considers the word sequence information and tries to preserve the semantic features. PV-DM takes text documents as inputs and outputs their vector representations in a latent semantic space. The visual feature $\bm{F}_{v,i}$ is extracted by the Caffe reference model~\cite{jia2014caffe}, which consists of 5 convolutional layers and 3 fully connected layers. This model was pre-trained on 1.2 million ImageNet (ILSVRC2010) images. In this work, we take the output of the second fully connected layer, resulting in a 4096-D feature vector as the visual features for each item.

After extracting the textual and visual features of items, we fuse them to better represent the characteristics of items~\cite{Wang2018FirstPersonDA}. Many multimodal feature fusion methods have been proposed and we adopt a widely used strategy\cite{zhang2014mm,srivastava2012multimodal} by first concatenating the textual and visual features and then feeding them into a multiple-layer neural networks. Specifically, the feature fusion network is,
\begin{gather}
\begin{split}
\bm{z}_{1} = \sigma( \bm{W}_{1}[\bm{F}_{t,i};\bm{F}_{v,i}] + \bm{b}_{1} ),\\
\bm{z}_{2} = \sigma(\bm{W}_{2}\bm{z}_{1} + \bm{b}_2), \\
......,~~~~~~~~~~~~~~~~~~~~~~~~~~~~~~~~~~~~~~~~~~~~~~~~~~~~~~~~~~~~~\\
 \bm{z}_{L}   = \sigma ( \bm{W}_{L} \bm{z}_{L-1} + \bm{b}_{L}  ),\\
\end{split}
\end{gather}
where $\bm{W}_{l}$ and $\bm{b}_{l}$ denote the weight matrix and bias vector for the $l$-th layer, respectively. $\sigma(\cdot)$ is the activation function and ReLU is adopted because its biologically plausible and  non-saturated property \cite{glorot2011deep}. The output of last layer is the fused feature, namely, $\bm{F}_{tv,i}=\bm{z}_{L}$. Noticed that our focus in this paper is to exploit items' multimodal features for capturing users' varying attentions on different aspects of various items.  The above extraction and fusion method is adopted for simplicity, and other advanced feature extraction and fusion methods could also be adopted in this case.

\subsubsection{Ranking loss weight} \label{sec:rlw}
We adopt the Weighted Approximate-Rank Pairwise (WARP) loss~\cite{weston2010large} to compute $\omega_{u,i}$. This scheme penalizes a positive item at a lower rank much more heavily than the one at the top, and produces the state-of-the-art results in previous work \cite{hsieh2017collaborative,zhao2015improving}. Given a metric $d$,  let  $rank_{d} ( u,i )$ denote the position of $i$ in $u$'s recommended ranking list, $\omega_{u,i}$ is computed as,
\begin{equation}
\omega_{u,i} = log( rank_{d} ( u,i ) + 1 ).
\end{equation}
In WARP, $ rank_{d} ( u,i )$ is estimated through a sequential sampling procedure which repeatedly samples  negative items to find impostors~\cite{weston2010large}. For each user-item pair $ (u,i)$, let $J$ denote the total number of items and $M$ denote the number of impostors in $N$ samples. The $rank_{d} ( u,i )$ is approximated as $\left \lfloor (J\times M)/N \right \rfloor$. For more details on WARP, please refer to~\cite{weston2010large}. In our implementation, we follow the procedure described in~\cite{hsieh2017collaborative} to estimate $rank_{d} ( u,i )$.

\subsubsection{Regularization.} \label{sec:reg}
As text reviews and item images represent items' characteristics, we would like that items with similar textual and visual features to be closer in the latent feature vector. To achieve the goal, we define the following $L_2$ loss function, 
\begin{equation}
L_{f}(q_*) =\sum_i \|\bm{F}_{tv,i}-\bm{q}_i\|.
\end{equation}
This function penalizes an item $i$'s feature vector $\bm{q}_i$ when $\bm{q}_i$  deviates away from the extracted feature $\bm{F}_{tv,i}$. 

To prevent the redundant of each dimension in the feature space,we then employ another regularization technique, covariance regularization~\cite{cogswell2015reducing} to reduce the correlation between the activation in a deep neural network. This technique can be also used in our model to de-correlate the dimensions in the feature space and thus to maximize the utilization of the given space. Let $\mathbf{y}^{n}$ denote the latent vector of an object, which could be a user or an item; and $n$ indexes the object in a batch of size $N$. The covariances between all pairs of dimensions $i$ and $j$ from a matrix C are defined as:
\begin{equation}
C_{i,j} = \frac{1}{N}\sum_{n}\left ( y_{i}^{n} - \mu _{i} \right )\left ( y_{j}^{n} - \mu _{j} \right ),
\end{equation}
where $\mu _{i} = \frac{1}{N}\sum_{n}y_{i}^{n}$. We define the loss $L_{c}$ to regularize the covariances:
\begin{equation}
L_{c} = \frac{1}{N}\left ( \left \| C \right \|_{f} - \left \| diag\left ( C \right ) \right \|_{2}^{2} \right ),
\end{equation}
where $\left \| \cdot \right \|_{f}$ is the Frobenius norm.

Finally, similar to the previous metric-based collaborative filtering methods~\cite{hsieh2017collaborative,tay2018latent}, we also bound all the user and item vector within a Euclidean unit sphere, i.e., $\| \bm{p}_{*}\|^{2} \leq 1$ and  $\| \bm{q}_{*} \|^{2} \leq 1$, for regularization and preventing overfitting.

\subsection{Optimization}
 With the consideration of all the regularization terms, the final object function of our MAML is,
 \begin{equation}
\begin{split}
\min \limits_{\mathbf{\theta,p_{*},q_{*}}} L_{m} + \lambda _{f}L_{f} + \lambda _{c}L_{c}\\
s.t. \quad \left \| \mathbf{p_{*}} \right \|^{2} \leq 1 \ and \ \left \| \mathbf{q_{*}} \right \|^{2} \leq 1 ,
\end{split}
\end{equation}
where $\lambda_{f}$ and $\lambda_{c}$ are hyperparameters that control the weight of each loss term. The optimization is quite standard and the stochastic gradient descent (SGD) algorithm is adopted. In implementation, the Adam optimizer~\cite{kingma2014adam} is adopted to tune the learning rate.

\section{EXPERIMENTS}

\subsection{Experimental Setup}
\subsubsection{Datasets}
The public Amazon review dataset~\cite{mcauley2013hidden}, which has been widely used for recommendation evaluation in previous studies ~\cite{zhang2017joint,chen2018attention,cheng2019mmalfm}, is adopted for experiments in this work. We adopted four product categories as shown in Table~\ref{tab:data}. Since we need to exploit reviews and item images to extract the item features, items without any textual reviews or images were removed. After this step, we further pre-processed the dataset to keep only the items and users which have at least 5 interactions.  The basic statistics of the four datasets are shown in Table~\ref{tab:data}. As we can see, the datasets are of different sizes and sparsity. For example, the \emph{office} dataset is relatively  denser than other datasets. The diversity of datasets is useful for analyzing the performance of our method and the competitors in different situations.

In this work, we focus on the top-$n$ recommendation task, which aims to recommend a set of $n$ top-ranked items that will be appealing to the target user. For each dataset, we randomly selected 70\% of the interactions from each user to construct the training set, and the remaining 30\% for testing. Each user has at least 5 interactions, thus we had at least 3 interactions per user for training, and at least 2 interactions per user for testing.

\subsubsection{Baselines and Evaluation Metrics}
We compare our MAML method  with the following baselines, including both shallow (BPR~\cite{rendle2009bpr}, VBPR~\cite{he2016vbpr}) and deep (DeepCoNN~\cite{zheng2017joint}, NeuCF~\cite{he2017neural}, JRL~\cite{zhang2017joint}) MF based models, as well as the metric learning based method CML~\cite{hsieh2017collaborative}. Besides, they cover different information sources, i.e.,  ratings (BPR, CML), reviews (DeepCoNN, CML$_F$~\cite{hsieh2017collaborative}, JRL), images (CML$_F$, JRL). Notice that CML$_F$ is an extension of CML. It extends the CML model by considering item features for top-$n$ recommendation. We use CML$_{text}$,  CML$_{image}$, and CML$_{all}$ to denote the model considering the text feature, image feature, and both features, respectively.

Four widely used evaluation metrics for top-$n$ recommendation are used in our evaluation, including \emph{precision}, \emph{recall}, \emph{NDCG}~\cite{jarvelin2002cumulated}, and \emph{hit ratio (HR)}.
All the metrics are computed based on the top 10 results. The reported results are the average values across all the testing users.
\begin{table}
	\caption{ Basic statistics of the experimental datasets.}
	\vspace{-0.2cm}
	\label{tab:data}
	\begin{tabular}{ccccl}
		\toprule
		Dataset&\#user&\#item&\#interactions&sparsity\\
		\midrule
		Office & 4,874& 2,406 & 53,137 & 99.55\%\\
		Men Clothing & 4,955& 5,028 & 32,363 &99.87\%\\
		Women Clothing & 19,244& 14,596 & 135,326 &99.95\%\\ 
		Toys Games & 18,748& 11,673 & 161,653&99.93\%\\
		\bottomrule
	\end{tabular}
	\vspace{-0.2cm}
\end{table}

\subsubsection{Experimental Settings}
We implemented our model with TensorFlow~\footnote{https://www.tensorflow.org.} and carefully tuned the key parameters. Specifically, we tuned the initial learning rate $\ell_0 \in \{0.1, 0.01, 0.001, 0.0001\}$; the margin $m$  in the hinge loss $ m \in \{0.1, 0.2, \dots, 2.0\}$; the number of negative samples ($s$) and the regularization parameters $\{s, \lambda_{f}, \lambda_{c}\} \in \{1, 2, \dots, 10\}$. In our experiments, the following setting works well: $\ell_0= 0.001$, $m$ =1.6 (1.5 for ToysGames), $s$ = 8 (4 for Office and MenClothing). $\lambda_{c}$ = 5, and $\lambda_{f}$ = 7 (2 for ToysGames). The optimal batch size varies across different datasets. Besides, model parameters are saved in every 10 epochs and the models are trained in maximum 1,000 epochs. In experiments, the dimension of latent vectors (i.e., $\bm{p}_u$ and $\bm{q}_i$) of all methods is set to 64. Notice that all the competitors have also been carefully tuned for fair comparisons. For example, for CML, we found that $\lambda_{c}$ = 1 (all the other hyperparameters are the same as our MAML model) works better. We released the codes and involved parameter settings to facilitate others to repeat this work\footnote{https://github.com/liufancs/MAML}.

\begin{table*}[ht]
	\caption{Performance of our MAML model and the competitors over four datasets. Noticed that the values are reported by percentage with '\%' omitted.} 
	\centering
	\resizebox{\textwidth}{33mm}{
		\begin{tabular}{l|cccc|cccc|cccc|cccc}
			\toprule
			Datasets&\multicolumn{4}{c|}{Office}&\multicolumn{4}{c|}{Men Clothing}&\multicolumn{4}{c|}{Women Clothing}&\multicolumn{4}{c}{Toys Games}\\
		    \cmidrule(r){1-5} \cmidrule(r){6-9} \cmidrule(r){10-13}\cmidrule(r){14-17}
			Metrics& NCDG & RECALL & HR & PR
			& NCDG & RECALL & HR & PR
			& NCDG & RECALL & HR & PR
			& NCDG & RECALL & HR & PR\\
			\midrule
			BPR &4.557&7.780&21.410&2.562&1.229&2.375&4.611&0.478&0.572&1.086&2.062&0.244&0.807&1.733&3.235&0.385\\
			NeuCF &4.145&6.609&19.504&2.516&1.416&2.599&6.017&0.625&1.136&2.115&4.377&0.491&2.374&4.558&9.801&1.374\\
			CML &6.680&8.986&24.025&3.141&2.520&4.040&7.815&0.811&3.026&4.542&9.157&1.034&6.207&8.638&17.454&2.204\\
			\midrule
			DeepCoNN &2.418&3.842&11.273&2.160&0.813&1.510&3.643&0.491&0.648&1.214&2.474&0.394&1.496&2.581&5.665&1.145\\
			CML$_{text}$ &6.889&9.171&24.778&3.223&3.212&4.952&9.659&1.005&3.507&5.139&9.972&1.144&6.345&8.787&17.957&2.260\\
			MAML$_{text}$ &7.094&9.346&25.165&3.222&3.282&4.961&9.843&1.019&3.590&5.246&10.329&1.175&6.410&8.954&18.093&2.289
			\\
			\midrule
			VBPR &3.072&4.942&15.381&1.817&1.061&1.914&3.873&0.392&0.921&1.644&3.357&0.377&0.846&1.595&3.738&0.418\\
			CML$_{image}$ &6.987&9.232&24.560&3.214&3.542&5.259&10.419&1.086&3.704&5.499&10.804&1.237&6.362&8.923&17.951&2.301
			\\
			MAML$_{image}$ &7.077&9.286&25.155&3.261&3.585&5.401&10.927&1.148&3.733&5.554&10.956&1.246&6.486&9.087&18.312&2.314\\
			\midrule
			JRL &3.391&3.421&12.345&1.525&2.574&3.920&7.247&0.657&1.729&2.906&4.965&0.457&2.302&4.599&8.232&1.382\\
			CML$_{all}$ &7.032&9.349&24.911&3.236&3.718&5.408&10.627&1.127&3.798&5.562&10.985&1.241&6.409&8.928&	18.135&2.308
			\\
			MAML$_{all}$ &\textbf{7.139*}&\textbf{9.386*}&\textbf{25.177*}&\textbf{3.265*}&\textbf{3.733*}&\textbf{5.574*}&\textbf{10.719*}&\textbf{1.152*}&\textbf{3.809*}&\textbf{5.636*}&\textbf{11.144*}&\textbf{1.260*}&\textbf{6.547*}&\textbf{9.111*}&\textbf{18.341*}&\textbf{2.327*}\\
			\bottomrule
		\end{tabular}}
		\begin{tablenotes}
		\footnotesize
		\item The symbol * denotes that the improvement is significant with $p-value < 0.05$ based on a two-tailed paired t-test.
	\end{tablenotes}
		\label{tab:results}
		\vspace{-0.3cm}
	\end{table*}

\subsection{Performance Comparison}
The results of our model and all the competitors  over the four datasets are reported in Table \ref{tab:results}.  It can be found that our method outperforms all the competitors consistently across all the test datasets in terms of different metrics. Besides, by grouping all the methods into four categories, we gained some interesting observations.

Firstly, we focused the performance of the methods in the first block, which only use the user-item interaction information. Specifically, NeuCF adopts neural networks which can better model user-item interactions and achieve better results than BPR in general. The \emph{Office} dataset has a smaller size but is relatively denser, which might be the reason that BPR can obtain good performance on this dataset.\footnote{Remind that MF based method suffers from the cold-start problem (i.e., very limited interactions for users or items) and NeuMF needs relatively more data for better learning.} CML outperforms both BPR and NeuMF by a large margin, because CML can capture fine-grained user preferences by using the metric-based learning approach. It can encode not only user preference but also the user-user and item-item similarity. 

The second and third blocks are the methods exploiting one type of side information, i.e., text and image, respectively. MAML$_{text}$ and MAML$_{image}$ are variants of our model MAML, which exploits only text and image, respectively. As we can see,  the recommendation performance can be greatly improved with the exploitation of additional item features, which has been demonstrated in many previous studies. DeepCoNN utilizes text information but performs unsatisfactorily. This is because the DeepCoNN uses text reviews not only in training but also in the test stage. In real scenario, the user review for an item is unavailable before the purchasing behavior happened. Therefore, in our experiments, we did not use review information in the testing stage, resulting in the performance degradation of DeepCoNN (which has also been demonstrated  in~\cite{catherine2017transnets}). It is interesting  that the performance based on images outperforms that based on texts (by comparing CML$_{image}$ and MAML$_{image}$ to  CML$_{text}$ and MAML$_{text}$). This might because the image features are more important for the selected four types of products (e.g., \emph{Clothing}).  VBPR achieves satisfactory performance in \emph{Women Clothing} but it is inferior to  NeuCF and BPR on other datasets. 

Finally, in the last block, the methods exploit both text and image features. The best performance of CML$_{all}$ and MAML$_{all}$ demonstrates the effectiveness of fusing multimodal features, which can indeed enhance the recommendation accuracy by considering multi-source item information. JRL achieves the best results over all the other baselines except the CML based ones (and BRP on the \emph{Office} dataset). The CML based methods outperform all the other baselines, verifying the potential of metric-based learning approach. The only difference between MAML and CML is that the former models user diverse preferences by using the attention mechanism. The consistent and significant improvement of MAML over CML validates the importance of considering user diverse preferences on different aspects of items, and also demonstrates the effectiveness of our proposed model.

\begin{figure}[]
		\centering 
		\includegraphics[width=0.965\linewidth]{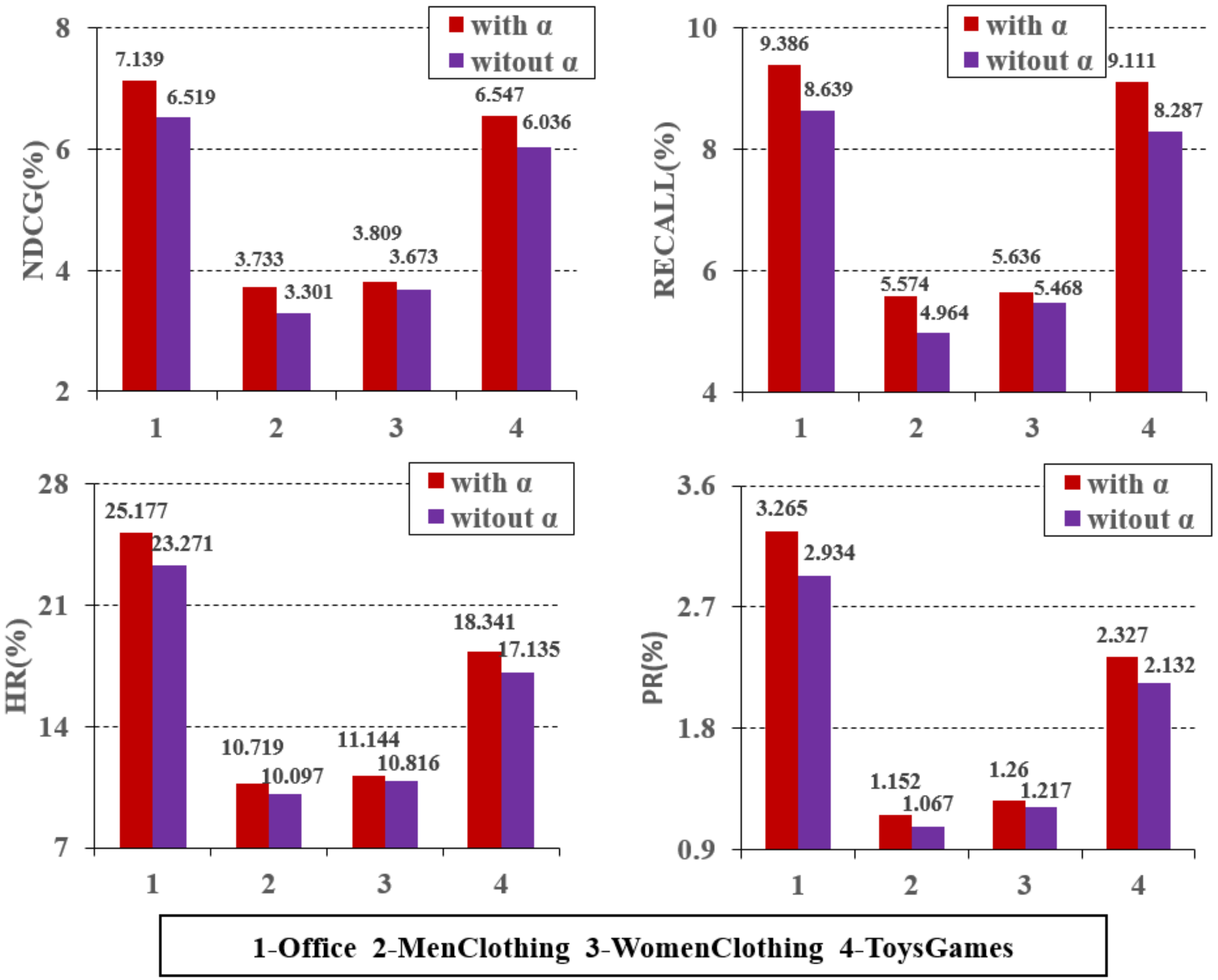} 
		\caption{Our adapted attention mechanism (with $\alpha$) \emph{vs.} the standard attention mechanism (whiteout $\alpha$).}  
		\label{fig:alphacompare}  
		\vspace{-0.5cm}
\end{figure}
\vspace{-0.01cm}
\subsection{Effects of Our Attention Mechanism} \label{alpha_section}
Remind that in our MAML model, we adapt the standard attention mechanism by enlarging the sum of all the attention weights by a factor $\alpha$. In our implementation, the value of $\alpha$ is set to be equal to the dimension of the user/item feature vectors, i.e., $\alpha = f$ (see Sect.~\ref{sec:am} for details). In this section, we examine the effectiveness of this tailored attention mechanism in our model. In particular, we compare the performance of our model with the adapted attention mechanism (with $\alpha$) and the standard one (without $\alpha$).  We tuned MAML with the standard attention mechanism carefully (e.g., $m\in\{0.001, 0.005, 0.01, \dots, 2.0 \}$, and $\{\lambda_{f}, \lambda_{c}\}\in \{0.0001, 0.001, \dots, 10\}$). The best results are used for comparisons. As shown in Fig.~\ref{fig:alphacompare}, with the adapted attention mechanism, our model can achieve consistent improvement by a large margin across all the test datasets, indicating the effectiveness of our attention mechanism in MAML.

\begin{figure}[]
    \centering
    \subfloat[user-items]{\label{fig:attentionleft}{\includegraphics[width=0.23\textwidth]{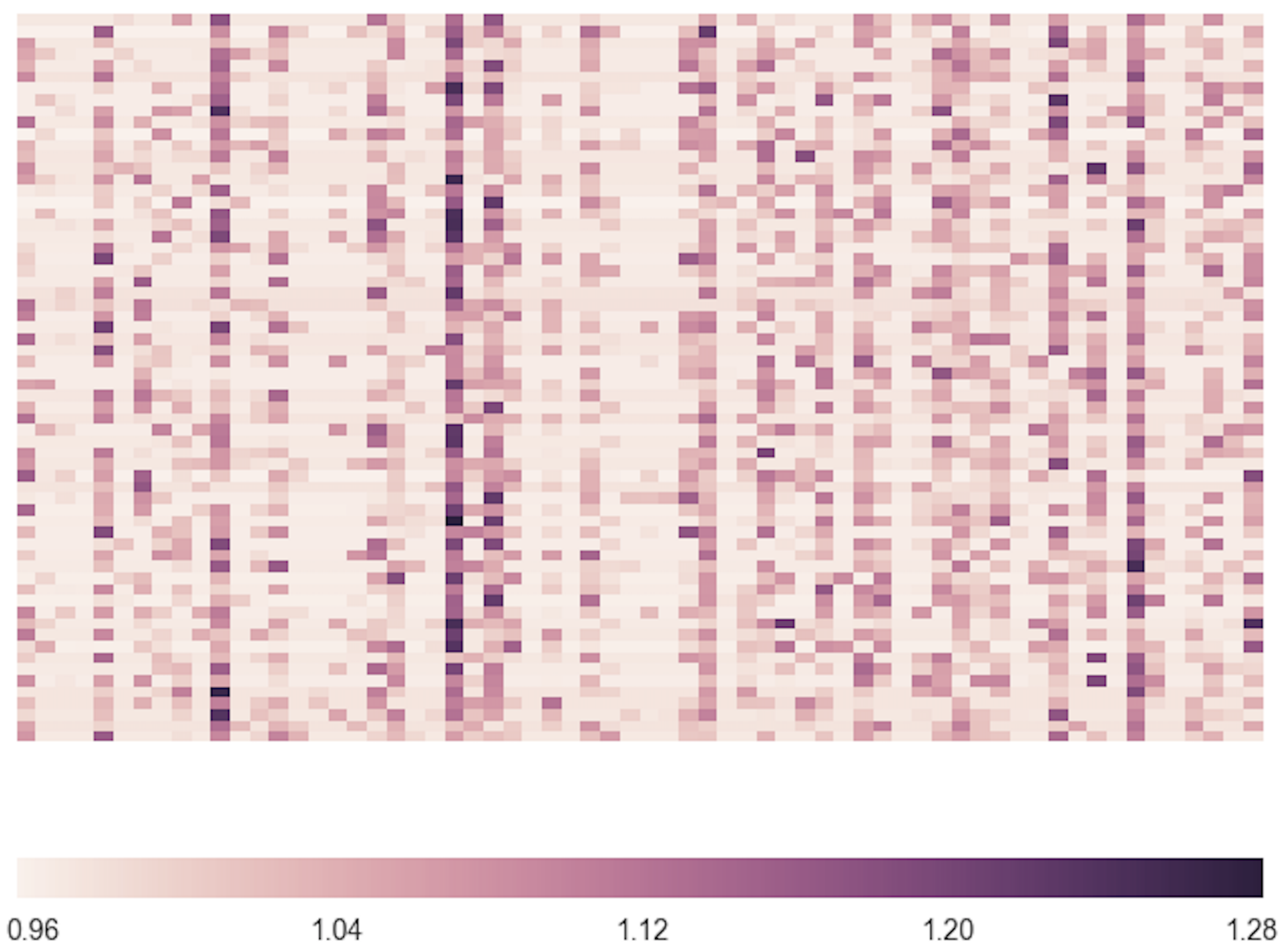}}}
    \subfloat[item-users]{\label{fig:attentionright}{\includegraphics[width=0.23\textwidth]{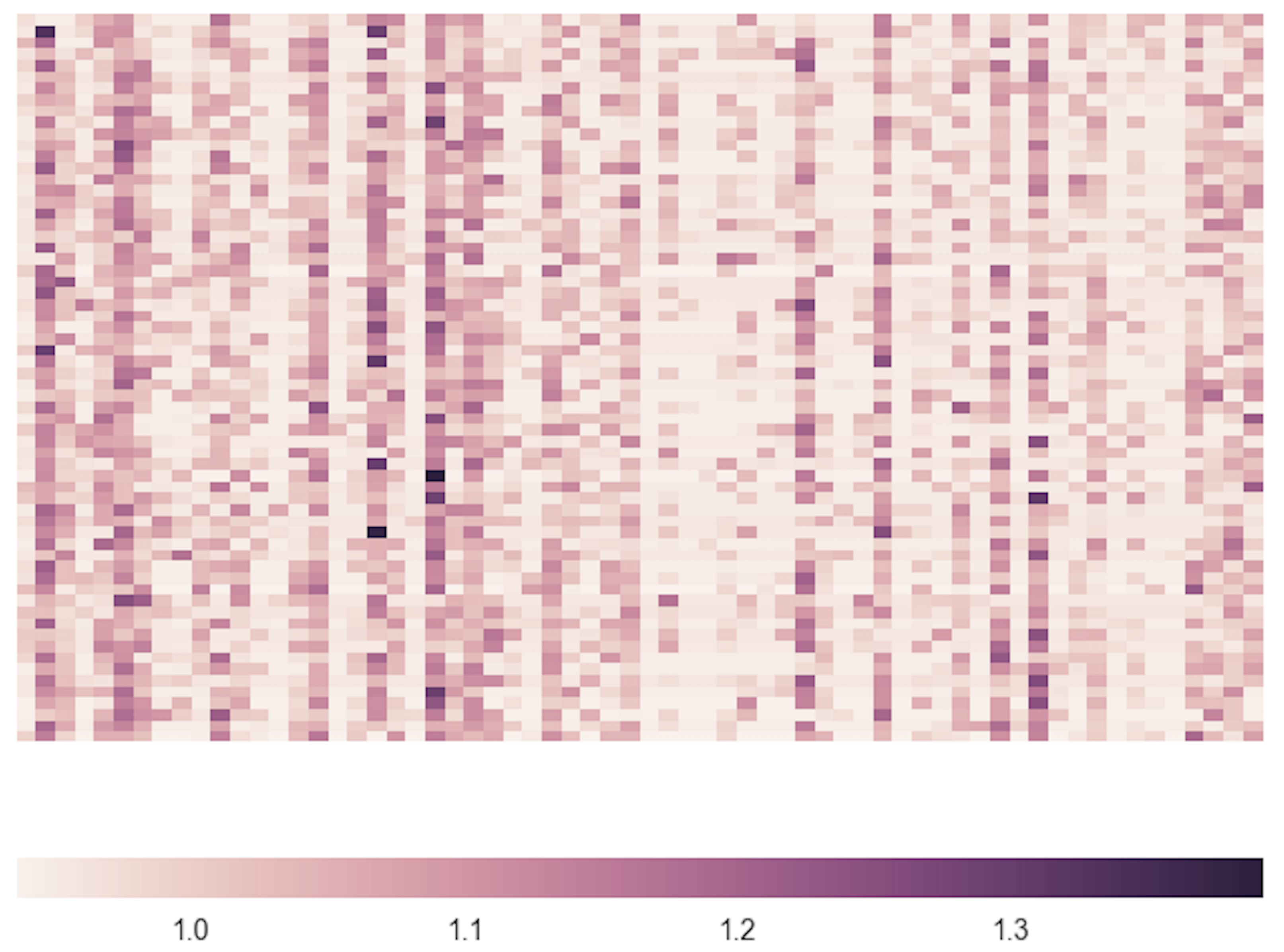}}}
    \caption{Visualization of Attention Vector.}
    \label{fig:attention}
    \vspace{-0.5cm}
\end{figure}

\subsection{Visualization}
In the proposed method, we model user diverse preferences by capturing user attention to different aspects of items. Our intuition is that the preference of a user on different aspects is varying when facing different items. Therefore, for each user-item pair, our model computes a unique attention weight for each aspect of the item (i.e., different dimensions of the feature vector). In this section, we would like to: 1) visualize the learned attention weights to illustrate user varying preferences on different aspects of various items; and 2) visualize user diverse preferences based on the purchased items.

\textbf{Attention Visualization}. We use heatmap to visualize the attention weights as shown in Fig.~\ref{fig:attention}. The color scale represents the intensities of attention weights, where a deeper color indicates a higher value and a lighter color indicates a lower value. Fig.~\ref{fig:attentionleft} visualizes the attention weights of a user for 64 items that were purchased by the user (in the \emph{Office} dataset). The $x$-axis represents the dimensions  of attention vector (i.e., 64), and $y$-axis represents the 64 items.  At the beginning,  the user attention on all aspects (or dimensions) is set to be the same (i.e., the all the values in the attention vector is 1). During the training process, the attention weight of each dimension will gradually adjust within the range of $[0, 64]$, and finally converge to a constant value. Fig.~\ref{fig:attentionleft} visualizes the converged values, which demonstrates that the user preference on different aspects is varying across different items. Fig.~\ref{fig:attentionright} illustrates the attention weights of 64 users on the aspects of a randomly selected item (also from the \emph{Office} dataset).
The $x$-axis also represents the dimensions of attention vector, and $y$-axis represents different users. We can see that, different users pay attention to different aspects of the same item.

\begin{figure}[]
    \centering
    \subfloat[Office]{\label{fig:clusterleft}{\includegraphics[width=0.44\textwidth]{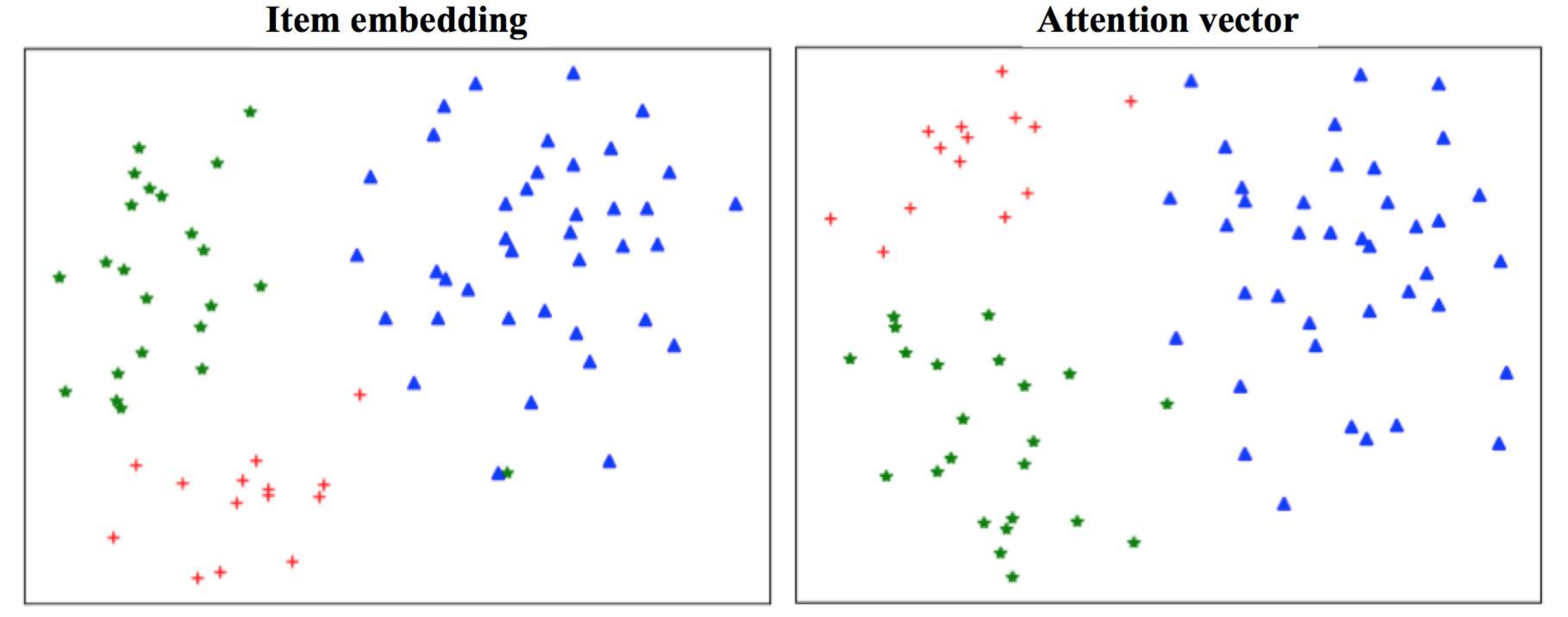}}}
    \vspace{-0.3cm}
    \subfloat[WomenClothing]{\label{fig:clusterright}{\includegraphics[width=0.44\textwidth]{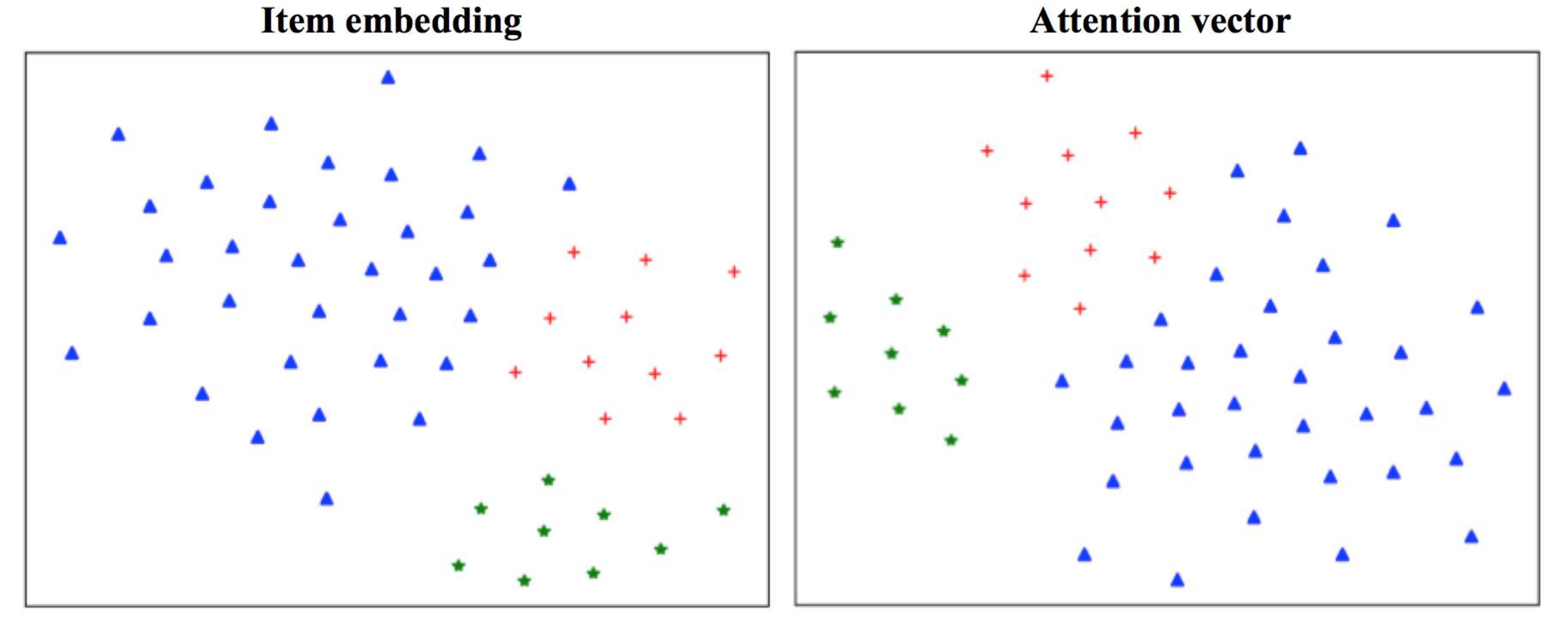}}}
    \caption{User preference visualization based on item embedding and attention vectors.}
    \label{fig:vup}
    \vspace{-0.3cm}
\end{figure}
\vspace{-0.01cm}
\textbf{User diverse preference visualization}.
We argued that a user preference is diverse, because a user 1) likes items with diverse features; and 2) has different preferences on the aspects of various items. In Fig.~\ref{fig:vup}, we visualize the user preference by clustering the purchased items using 1) item embedding and 2) user attention vectors with respect to items. Fig.~\ref{fig:vup} shows the clustering results of all the items that a user purchased from \emph{Office} and \emph{WomenClothing}, respectively. The left figures visualize the clustering results based on item embedding (i.e., $\bm{q}_i$); and the right ones illustrate the clustering results based on user attention vectors. Every dot in the figures denotes an item. Besides, in both Fig.~\ref{fig:clusterleft} and Fig.~\ref{fig:clusterright}, the dots with the same color (e.g., blue) in the two figures denote the same set of items. We use t-SNE~\cite{vanDerMaaten2008} for clustering and visualization. Notice that the item vector (i.e., $\bm{q}_i$)  is learned based on its interactions with all the users. Therefore, this vector should well characterize all the aspects of the item. In contrast, an attention vector $\bm{a}_{u,i}$ characterizes a user's attention to the different aspects of an item. Therefore, the clustering results  based on the item embedding (left figures) can demonstrate the user diverse preferences on items with different features; the clustering results based on the attention vector (right figures) validate user varying attention on different aspects of items. Besides, we can see that the distribution of the same items (e.g., red dots in Fig.~\ref{fig:clusterleft}) is different in the two types of figures. It  also demonstrates that a user pays different attention to the various aspects of different items. 

\section{CONCLUSIONS}
In this paper, we presented a novel Multimodal Attentive Metric Learning (MAML) model for top-n recommendation. In particular, this model is designed to model user diverse preferences on different aspects of items. This is achieved by a proposed attention neural network, which exploits the multimodal features of the target item to estimate a user's specific attention on each aspect of this item. Moreover, our model is developed based on a metric-based learning approach, which avoids the inherent limitation of matrix factorization based methods and thus can capture fine-grained user preference. Extensive experiments have been conducted on four real-world datasets from Amazon. The results show that our method significantly outperforms a variety of competitors, demonstrating the effectiveness of our method and the potential of modeling user diverse preference for recommendation. Besides, we also studied the effectiveness of the designed attention mechanism and visualized the user diverse preference based on both the learned item vectors and the attention vectors. 

\section{Acknowledgments}
This work is supported by the National Natural Science Foundation of China, No.:61772310, No.:61702300, No.:61702302, No.:61802231, No.:U1836216; the Project of Thousand Youth Talents 2016; the Shandong Provincial Natural Science and Foundation, No.:ZR2019JQ23, No.:ZR2019QF001; the Future Talents Research Funds of Shandong University, No.:2018WLJH63.

%
\newpage
\bibliographystyle{ACM-Reference-Format}
\bibliography{ms}

%









\end{document}